\documentclass{article}
\usepackage[utf8]{inputenc}
\usepackage[affil-it]{authblk}
\usepackage{hyperref}
\usepackage{bm}
\usepackage{float}
\usepackage{amsmath}
\usepackage{amssymb}
\usepackage{graphicx}
\usepackage{rotating}
\usepackage{caption}
\usepackage{comment}
\usepackage{xcolor}
\usepackage[left=4cm, right=4cm]{geometry}
\usepackage[colorinlistoftodos]{todonotes}
\usepackage[tableposition=top]{caption}

\title{Distributed Representations of Atoms and Materials for Machine Learning}
\author[1]{Luis M. Antunes} 
\author[1]{Ricardo Grau-Crespo}
\author[1,2]{Keith T. Butler}
\date{}
\affil[1]{\footnotesize Department of Chemistry, University of Reading, Whiteknights, Reading RG6 6DX, United Kingdom. {\normalfont l.m.antunes@pgr.reading.ac.uk}}
\affil[2]{\footnotesize SciML, Scientific Computing Department, Rutherford Appleton Laboratory, Harwell OX11 0QX, United Kingdom.}

\begin{document}

\maketitle

\begin{abstract}
    The use of machine learning is becoming increasingly common in computational materials science. To build effective models of the chemistry of materials, useful machine-based representations of atoms and their compounds are required. We derive distributed representations of compounds from their chemical formulas only, via pooling operations of distributed representations of atoms. These compound representations are evaluated on ten different tasks, such as the prediction of formation energy and band gap, and are found to be competitive with existing benchmarks that make use of structure, and even superior in cases where only composition is available. Finally, we introduce a new approach for learning distributed representations of atoms, named SkipAtom, which makes use of the growing information in materials structure databases.
\end{abstract}

\section{Introduction}

In recent years, the study of machine learning (ML) has had a significant impact on many disciplines. Accordingly, materials science and chemistry has recently seen a surge in interest in applying the most recent advances in ML to the problems of the field  \cite{himanen2019data, schleder2019dft, goldsmith2018machine, butler2018machine, ramprasad2017machine, schmidt2019recent}. A central problem in materials science is the rational design of materials with specific properties. Typically, useful materials have been discovered serendipitously \cite{disalvo2000challenges}. With the advent of ubiquitous and capable computing infrastructure, materials discovery has been increasingly aided by computational chemistry, especially density functional theory (DFT) simulations \cite{zunger2018inverse}. Such theoretical calculations are indispensable when investigating the properties of novel materials. However, they are computationally intensive, and performing such analysis on large numbers of compounds (there are more than $10^{10}$ chemically sensible stoichiometric quaternary compounds possible \cite{davies2016computational}) becomes impractical with today's computing technology. Moreover, certain chemical systems, such as those with very strongly correlated electrons, or with high levels of disorder, remain a theoretical challenge to DFT \cite{duan2021putting, midgley2021bandgap}. 

The application of ML to materials science aims to ameliorate some of these problems, by providing alternate computational routes to properties of interest. There have been numerous examples of the successful application of ML to chemical systems. Techniques from ML have been used to predict very local and detailed properties, such as atomic and molecular orbital energies and geometries \cite{qiao2020orbnet} or partial charges\cite{raza2020message}, and also global properties, such as the formation energy and band gap of a given compound \cite{faber2015crystal, zhuo2018predicting, davies2019data, artrith2019machine}.

For a ML algorithm to work effectively, the objects of the system of interest must be converted into faithful representations that can be consumed in a computational context. Deriving such representations has been a main focus for researchers in ML, and in the case of Deep Learning, such representations are typically learned automatically, as part of the training process \cite{lecun2015deep}. Related to this is the concept of Unsupervised Learning, where patterns in the data are derived without the use of labels, or other forms of supervision \cite{hinton1999unsupervised}. Indeed, given that most data is unlabelled, such techniques are very valuable. Some of the most successful and widely used algorithms, such as Word2Vec from the field of Natural Language Processing (NLP), use unsupervised learning to derive effective representations of the objects in the system of interest (words, in this case) \cite{mikolov2013efficient, le2014distributed}.

The most basic object of interest in chemical systems is very often the atom. Thus, there have already been several investigations examining the derivation of effective machine representations of atoms in an unsupervised setting  \cite{zhou2018learning, tshitoyan2019unsupervised, chakravarti2018distributed}, and other investigations have aimed to learn good atomic representations in the context of a supervised learning task. \cite{jha2018elemnet, goodall2020predicting} A learned representation of an atom generally takes the form of an embedding, which can be described as a relatively low-dimensional space in which higher-dimensional vectors can be expressed. Using embeddings in a ML task is advantageous, as the number of input dimensions is typically lower than if higher-dimensional sparse vectors were used. Moreover, embeddings which are semantically similar reside closer together in space, which provides a more principled structure to the input data. Such representations should allow ML models to learn a task more quickly and effectively.

A widely held hypothesis in ML is that unlabelled data can be used to learn effective representations. In this work, we introduce a new approach for learning atomic representations using an unsupervised approach. This approach, which we name SkipAtom, is inspired by the Skip-gram model in NLP, and takes advantage of the large number of inorganic structures in materials databases. We also investigate forming representations of chemical compounds by pooling atomic representations. Combining vectors by various pooling operations to create representations of systems composed from parts (e.g. sentences from words) is a common technique in NLP, but apparently remains largely unexplored in materials informatics \cite{mitchell2008vector}. The analogy we explore here is that atoms are to compounds as words are to sentences, and our results demonstrate that effective representations of compounds can be composed from the vector representations of the constituent atoms. Finally, a common problem when searching chemical space for new materials is that the structure of a compound may not be known. Since the properties of a material are typically tightly coupled to its structure, this creates a significant barrier \cite{meredig2014combinatorial}. Here, we compare our models, which operate on representations derived from chemical formulas only, to benchmarks that are based on models that use structural information. We find that, for certain tasks, the performance of the composition-only models is comparable.

\section{Representations of Atoms and Compounds}

There are various strategies for providing an atom with a machine representation. These range from very simple and unstructured approaches, such as assigning a random vector to each atom, to more sophisticated approaches, such as learning distributed representations. A distributed representation is a characterization of an object attained by embedding in a continuous vector space, such that similar objects will be closer together. 

Similarly, a compound may be assigned a machine representation. Again, these representations may be learned on a case-by-case basis, or they may be formed by composing existing representations of the corresponding atoms.

\subsection{Atomic Representations}

We are interested in deriving representations of atoms that can be used in a computational context, such as a ML task. Intuitively, we would like the representations of similar atoms to be similar as well. Given that atoms are multifaceted objects, a natural choice for a computational descriptor for an atom might be a vector: an $n$-tuple of real numbers. Vector spaces are well understood, and can provide the degrees of freedom necessary to express the various facets that constitute an atom. Moreover, with an appropriately selected vector space, such atomic representations can be subjected to the various vector operations to quantify relationships and to compose descriptions of systems of atoms, or compounds.

\subsubsection{Random Vectors}

The simplest approach to assigning a vector description to an atom is to simply draw a random vector from $\mathbb{R}^n$, and assign it to the atom. Such vectors can come from any distribution desired, but in this report, such vectors will come from the standard normal distribution, $\mathcal{N}(0,1)$.

\subsubsection{One-hot Vectors}

\begin{figure}[H]
\begin{center}
\includegraphics[scale=0.45]{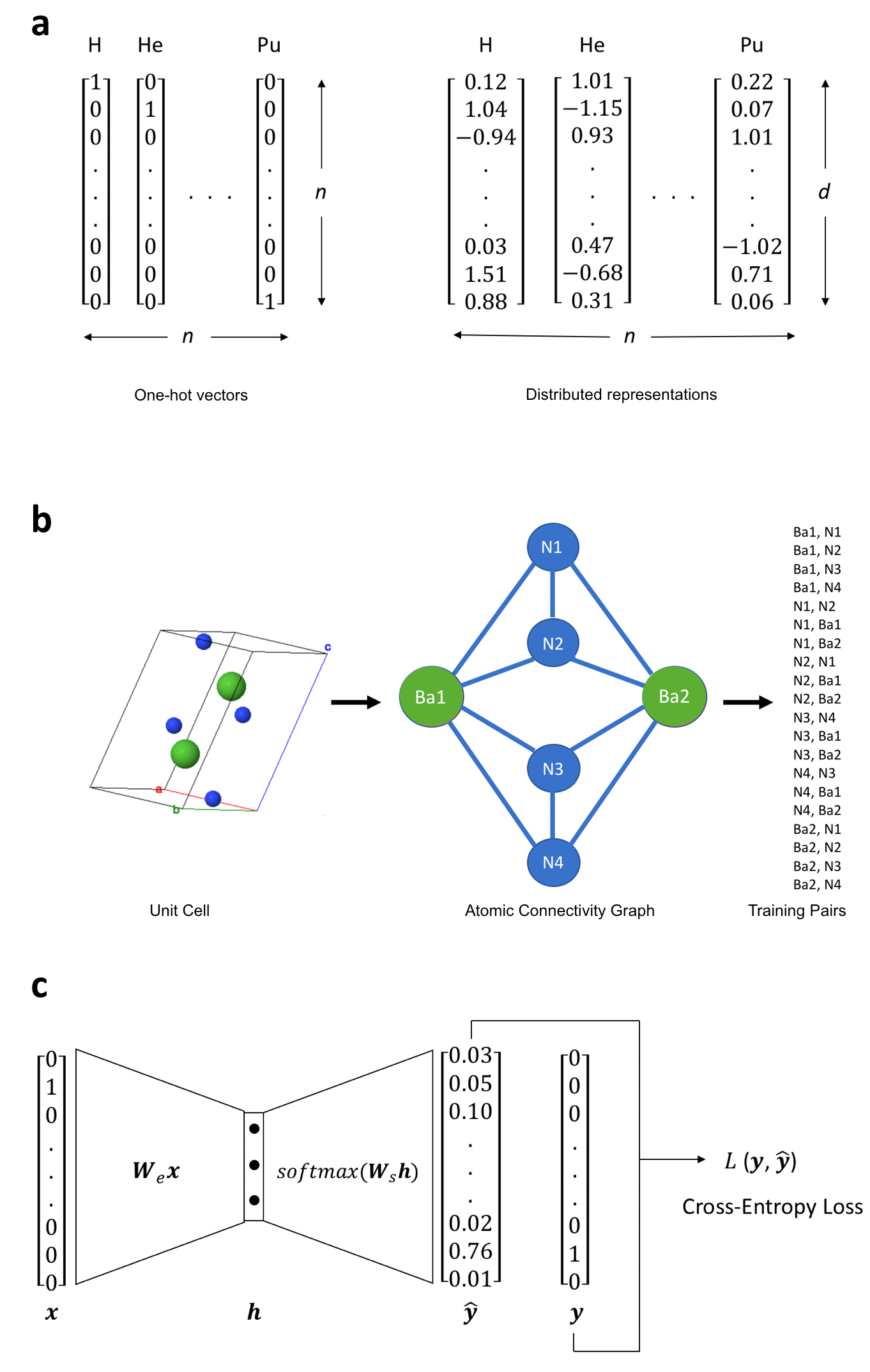}
\end{center}
\caption{\textbf{a} Scheme illustrating one-hot and distributed representations of atoms. In the diagram, there are $n$ atoms represented, and $d$ is the adjustable number of dimensions of the distributed representation. Note that the atoms in this example are H, He and Pu, but they could be any atom. \textbf{b} Scheme describing how training data is derived for the creation of SkipAtom vectors. Here, a graph representing the atomic connectivity in the structure of Ba\textsubscript{2}N\textsubscript{4} is depicted, and the resulting target-context atom pairs derived for training. The graph is derived from the unit cell of Ba\textsubscript{2}N\textsubscript{4}. \textbf{c} Scheme describing how the SkipAtom vectors are derived through training. Here, a one-hot vector, $\pmb x$, representing a particular atom is transformed into an intermediate vector $\pmb h$ via multiplication with matrix $\pmb W_e$. The matrix $\pmb W_e$ is the embedding matrix, whose columns will be the final atom vectors after training. Training consists of minimizing the cross-entropy loss between the output vector $\bm{\hat{y}}$ and the one-hot vector representing the context atom, $\pmb y$. The output $\bm{\hat{y}}$ is obtained by applying the $softmax$ function to the product $\pmb W_{s}\pmb h$.}
\label{fig:skipatomcombined}
\end{figure}

One-hot vectors, common in ML, are binary vectors that are used for distinguishing between various categories.   One assigns a vector component to each category of interest, and sets the value of the corresponding component to 1 when the vector is describing a given category, and the value of all other components to 0. More formally, a one-hot $n$-dimensional vector $\pmb v$ is in the set $\{0, 1\}^n$ such that $\sum_{i=1}^{n} v_i = 1$, where $v_i$ is a component of $\pmb v$. A unique one-hot vector is assigned to each category. In the context of this report, a category is an atom (Figure \ref{fig:skipatomcombined}a).

\subsubsection{Atom2Vec}

If one may know a word by the company it keeps, then the same might be said of an atom. In 2018, Zhou and coworkers described an approach for deriving distributed atom vectors that involves generating a co-occurrence count matrix of atoms and their chemical environments, using an existing database of materials, and applying Singular Value Decomposition (SVD) to the matrix. \cite{zhou2018learning} The number of dimensions of the resulting atomic vectors is limited to the number of atoms used in the matrix. 

\subsubsection{Mat2Vec}

A popular means of generating word vectors in NLP is through the application of the Word2Vec algorithm, wherein an unsupervised learning task is employed \cite{mikolov2013efficient}. Given a corpus (a collection of text), the goal is to predict the likelihood of a word occurring in the context of another. A neural network architecture is employed, and the learned parameters of the projection layer constitute the word vectors that result after training. In 2019, Tshitoyan and coworkers described an approach for deriving distributed atom vectors by making direct use of the materials science literature \cite{tshitoyan2019unsupervised}. Instead of using a database of materials, they assembled a textual corpus from millions of scientific abstracts related to materials science research, and then applied the Word2Vec algorithm to derive the atom representations.

\subsubsection{SkipAtom}

In the NLP Skip-gram model, an occurrence of a word in a corpus is associated with the words that co-occur within a context window of a certain size. The task is to predict the context words given the target word. Although the aim is not to build a classifier, the act of tuning the parameters of the model so that it is able to predict the context of a word results in a parameter matrix that acts effectively as the embedding table for the words in the corpus. Words that share the same contexts should share similar semantic content, and this is reflected in the resulting learned low-dimensional space. Analogously, atoms that share the same chemo-structural environments should share similar chemistry.

In the SkipAtom approach, the crystal structures of materials from a database are used in the form of a graph, representing the local atomic connectivity in the material, to derive a dataset of connected atom pairs (Figure \ref{fig:skipatomcombined}b). Then, similarly to the Skip-gram approach of the Word2Vec algorithm, Maximum Likelihood Estimation is applied to the dataset to learn a model that aims to predict a context atom given a target atom.

More formally, a materials database consists of a set of materials, $M$. A material, $m \in M$, can be represented as an undirected graph, consisting of a set of atoms, $A_m$, comprising the material, and bonds $B_m  \subseteq \{(x,y) \in A_m \times A_m | x \neq y\}$, which are unordered pairs of atoms. The task is to maximize the average log probability:
\begin{equation} \label{eq:skipatom)}
\dfrac{1}{|M|} \sum_{m \in M} \sum_{a \in A_m} \sum_{n \in N(a)} \log{p(n|a)}
\end{equation}
where $N(a)$ are the neighbours of $a$ (not including $a$ itself); more specifically: 
$N(a) = \{x \in A_m | (a, x) \in B_m\}$.

In practice, this means that the cross-entropy loss between the one-hot vector representing the context atom and the normalized probabilities produced by the model, given the one-hot vector representing the target atom, is minimised (Figure \ref{fig:skipatomcombined}c).

The graph representing a material can be derived using any approach desired, but in this work, an approach is used which is based on a Voronoi algorithm, which identifies nearest neighbours using solid angle weights to determine the probability of various coordination environments.

The result of SkipAtom training is a set of vectors, one for each atom of interest (Figure \ref{fig:skipatomcombined}a), that reflects the unique chemical nature of the represented atom, as well as its relationship to other atoms.

A complicating factor in the procedure just described is that some atoms may be under-represented in the database, relative to others. This will result in the parameters of those infrequently occurring atoms receiving fewer updates during training, resulting in lower quality representations for those atoms. This is an issue when learning word representations as well, and there have been several solutions proposed in the context of NLP \cite{pilehvar2016conflated, pilehvar2017inducing}. Borrowing from these solutions, we apply an additional, optional processing step to the learned vectors, termed \textit{induction}. The aim is to adjust the learned vectors so that they reside in a more sensible area of the representation space. To achieve this, each atom is first represented as a triple, given by its periodic table group number and row number, and its electronegativity. Then, for each atom, the closest atoms are obtained, in terms of the cosine similarity between the vectors formed from these triples. Using the learned embeddings for these closest atoms, a \textit{mean nearest-neighbour representation} is derived, and the induced atom vector, $\hat{\pmb u}$, is formed by adding the original atom vector, $\pmb u$, to the mean nearest neighbour:
\begin{equation} \label{eq:skipatominduction)}
\hat{\pmb u} = \pmb u + \dfrac{1}{N} \sum_{i=0}^{N} e^{-i} \pmb v_i
\end{equation}
where $N$ is the number of closest atoms to consider, and $\pmb v_i$ is the learned embedding of the $i^{th}$ nearest atom from the sorted list of nearest atoms. In this work, the nearest 5 atoms are considered.

\subsection{Compound Representations}

Atom vectors by themselves may not be directly useful, as most problems in materials informatics involve chemical compounds. However, atom vectors can be combined to form representations of compounds.

\subsubsection{Atom Vector Pooling}

The most basic and general way of combining atom vectors to form a representation for a compound is to perform a pooling operation on the atom vectors corresponding to the atoms in the chemical formula for the compound. There are three common pooling operations: sum-pooling, mean-pooling, and max-pooling.

\textit{Sum-pooling} involves performing component-wise addition of the atom vectors for the atoms in the chemical formula. That is, for a chemical compound whose formula is comprised of $m$ constituent elements, and a set of atom vectors, $\pmb v \in V$, the compound vector, $\pmb u$, is given by:
\begin{equation} \label{eq:sumpooling)}
\sum_{i=1}^{m} c_i \pmb v_i
\end{equation}
where $\pmb v_i$ is the corresponding atom vector for the $i$th constituent element in the formula, and $c_i$ is the relative number of atoms of the $i$th constituent element (which need not be a whole number, as in the case of non-stoichiometric compounds).

\textit{Mean-pooling} involves performing component-wise addition of the atom vectors for the atoms in the chemical formula, followed by dividing by the total number of atoms in the formula:
\begin{equation} \label{eq:meanpooling)}
\frac{\sum_{i=1}^{m} c_i \pmb v_i}{\sum_{i=1}^{m} c_i}
\end{equation}

Finally, \textit{max-pooling} involves taking the maximum value for each component of the vectors being pooled:
\begin{equation} \label{eq:maxpooling)}
\max_{i=1}^{m} c_i \pmb v_i
\end{equation}
where $\max$ returns a vector where each component has the maximum value of that component across $n$ input vectors.

\subsubsection{ElemNet (Mean-pooled One-hot Vectors)}

If we assign a unique one-hot vector to each atom, and perform mean-pooling of these vectors when forming a representation for a chemical compound, then the result is the same as the input representation for the ElemNet model \cite{jha2018elemnet}. Such a compound vector is sparse (as most compounds do not typically contain more than 5 or 6 atom types). Each component of the vector contains the unit normalized amount of the atom in the formula. For example, for H\textsubscript{2}O, the component corresponding to H would have a value of 0.66 whereas the component corresponding to O would have a value of 0.33, and all other components would have a value of zero.

\subsubsection{Bag-of-Atoms (Sum-pooled One-hot Vectors)}

In NLP, the Bag-of-Words is a common representation used for sentences and documents. It is formed by simply performing sum-pooling of the one-hot vectors for each word in the text. Similarly, we can conceive of a Bag-of-Atoms representation for chemical informatics, where sum-pooling is performed with the one-hot vectors for the atoms in a chemical formula. The result is a list of counts of each atom type in the formula. This is an unscaled version of the ElemNet representation. Crucially, this sum-pooling of one-hot vectors is more appropriate for describing compounds than it is for describing natural language sentences, as there is no significance to the order of atoms in a chemical formula as there is for the order of words in a sentence.

\section{Experiments}

\subsection{Tasks}

A number of diverse materials ML tasks are utilized to evaluate the effectiveness of the pooled atom vector representations, and the quality of the SkipAtom representation. In total, ten previously described tasks are utilized, and are broadly divided into two categories: those used for evaluating the pooling approach, and those used for evaluating the SkipAtom approach. To evaluate the pooling approach, nine tasks are chosen, and are described in Table \ref{tab:tasktypes}.

\begin{table}[H]
\caption {The predictive tasks utilized in this study to evaluate the atom vector pooling approach. All datasets and benchmarks for the tasks above are described in \cite{dunn2020benchmarking}, with the exception of the Formation Energy task, which is described in \cite{jha2018elemnet}.}
\footnotesize
\begin{tabular*}{\textwidth}{ c c c c c }
\hline
\bf Task & \bf Type & \bf Examples & \bf Structure? & \bf Method \\
\hline
Band Gap (eV) & Regression & 4,604 & No & Experiment \cite{zhuo2018predicting} \\
Band Gap (eV) & Regression & 106,113 & Yes & DFT-GGA \cite{jain2013commentary, ong2015materials} \\
Bulk Modulus (log(GPa)) & Regression & 10,987 & Yes & DFT-GGA \cite{de2015charting} \\
Shear Modulus (log(GPa)) & Regression & 10,987 & Yes & DFT-GGA \cite{de2015charting} \\
Refractive Index ($n$) & Regression & 4,764 & Yes & DFPT-GGA \cite{petousis2017high} \\
Formation Energy (eV/atom) & Regression & 275,424 & Yes & DFT \cite{jha2018elemnet, saal2013materials} \\
Bulk Metallic Glass Formation & Classification & 5,680 & No & Experiment \cite{ward2016general, kawazoe1997nonequilibrium} \\
Metallicity & Classification & 4,921 & No & Experiment \cite{zhuo2018predicting} \\
Metallicity & Classification & 106,113 & Yes & DFT-GGA \cite{jain2013commentary, ong2015materials} \\
\hline
\end{tabular*}
\label{tab:tasktypes}
\end{table}

The tasks were chosen to represent the various scenarios encountered in materials data science, such as the availability of both smaller and larger datasets, the need for either regression or classification, the availability of material structure information, and the means (experiment or theory) by which the data is obtained. The OQMD (Open Quantum Materials Database) Formation Energy task \cite{jha2018elemnet, saal2013materials} requires a different training protocol, as it was derived from a different study than the other eight tasks that are used for the pooling approach, which were sourced from the Matbench test suite \cite{dunn2020benchmarking}. 

To evaluate the SkipAtom representation, the Elpasolite Formation Energy task was utilized. The task and the model were initially described in the paper that introduced Atom2Vec (an alternative approach for learning atom vectors) \cite{zhou2018learning}. The task consists of predicting the formation energy of elpasolites, which are comprised of a quaternary crystal structure, and have the general formula ABC\textsubscript{2}D\textsubscript{6}. The target formation energies for 5,645 examples were obtained by DFT \cite{faber2016machine}. The input consists of a concatenated sequence of atom vectors, each representing the A, B, C, and D atoms. We reproduce the approach here, for comparison against the Atom2Vec results. 

All tasks require a representation of a material as input, and produce a prediction of a physical property as output, in either a regression or classification setting. Moreover, with the exception of the Elpasolite Formation Energy task, all tasks make use of the same model architecture (described in detail below).

\subsection{Protocols}

For the purposes of evaluation, the atom and compound vectors were utilized as inputs to feed-forward neural networks. All results for evaluating the pooling approach were obtained using a 17-layer feed-forward neural network architecture based on ElemNet \cite{jha2018elemnet}. The network was comprised of 4 layers with 1,024 neurons, followed by 3 layers with 512 neurons, 3 layers with 256 neurons, 3 layers with 128 neurons, 2 layers with 64 neurons, and 1 layer with 32 neurons, all with ReLU activation. For regression tasks, the output layer consisted of a single neuron and linear activation. For classification tasks, the output layer consisted of a single neuron and sigmoid activation (as only binary classification was performed). Instead of using dropout layers for regularization, as in the ElemNet approach, L2 regularization was used, with a regularization constant of $10^{-5}$. The goal during training was to minimise the Mean Absolute Error loss (for regression tasks), or the Binary Cross-entropy loss (for classification tasks). All pooling approach experiments utilized a mini-batch size of 32, and a learning rate of $10^{-4}$ along with the Adam optimizer (with an epsilon parameter of $10^{-8}$) \cite{kingma2014adam}. As described in the paper that introduces the Matbench test set \cite{dunn2020benchmarking}, $k$-fold cross-validation was performed to evaluate the compound vectors in regression tasks, with the same random seed to ensure the same splits were used each time. For classification tasks, stratified $k$-fold cross-validation was performed. As required by the benchmarking protocol, 5 splits were used (with the exception of the OQMD Formation Energy prediction task, which used 10 splits). Because the variance was high for some tasks after $k$-fold cross-validation, repeated $k$-fold cross-validation was performed, to reduce the variance \cite{moss2018using}. All training was carried out for 100 epochs, and the best performing epoch was chosen as the result for that split. By following this protocol, a direct and fair comparison can be made to results reported previously using the same Matbench test set \cite{dunn2020benchmarking}.

The results for evaluating the SkipAtom approach were obtained using the Elpasolite neural network architecture and protocol, originally described in the paper that introduces Atom2Vec \cite{zhou2018learning}. The input to the neural network is a vector constructed by concatenating 4 atom vectors, representing each of the 4 atoms in an Elpasolite composition. The single hidden layer consists of 10 neurons, with ReLU activation. The output layer consists of a single neuron, with linear activation. L2 regularization was used, with a regularization constant of $10^{-5}$. The goal during training was to minimise the Mean Absolute Error loss. The training protocol differs slightly in this report, and 10-fold cross-validation was performed, utilizing the result after 200 epochs of training. The same random seed was used for all experiments, to ensure the same splits were utilized. A mini-batch size of 32 was utilized, and a learning rate of $10^{-3}$ along with the Adam optimizer (with an epsilon parameter of $10^{-8}$) was chosen \cite{kingma2014adam}.

Learning of the SkipAtom vectors involved the use of the Materials Project database \cite{Jain2013}. To assemble the training set, 126,335 inorganic compound structures were downloaded from the database. Each of these structures was converted into a graph representation using an approach based on a Voronoi algorithm, and a dataset of co-occurring atom pairs was derived. A total of 15,360,652 atom pairs were generated, utilizing 86 distinct atom types. The architecture consisted of a single hidden layer with linear activation, whose size depended on the desired dimensionality of the learned embeddings, and an output layer with 86 neurons (one for each of the utilized atom types) with $softmax$ activation. The training objective consisted of minimizing the cross-entropy loss between the predicted context atom probabilities and the one-hot vector representing the context atom, given the one-vector representing the target atom as input. Training utilized stochastic gradient descent with the Adam optimizer, with a learning rate of $10^{-2}$ and a mini-batch size of 1,024, for 10 epochs.

\subsection{Results and Discussion}

\begin{figure}[H]
\begin{center}
\includegraphics[scale=0.33]{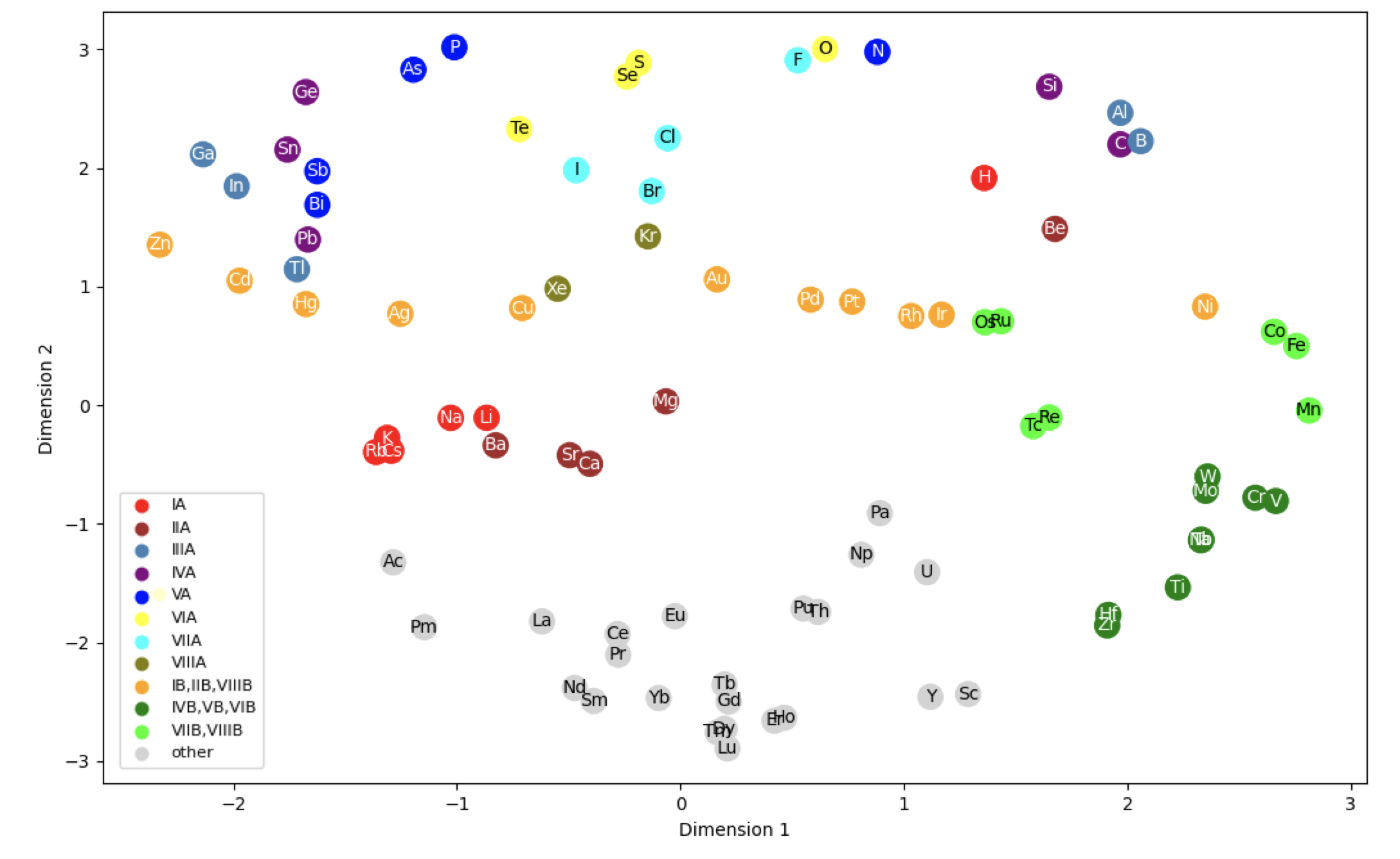}
\end{center}
\caption{Dimensionally reduced SkipAtom atom vectors with an original size of 200 dimensions. The vectors were reduced to 2 dimensions using t-SNE.}
\label{fig:skipatomreduced}
\end{figure}

A common technique for making high-dimensional data easier to visualize is t-SNE (t-Stochastic Neighbour Embedding) \cite{van2008visualizing}. Such a technique reduces the dimensionality of the data, typically to 2 dimensions, so that it can be plotted. Visualizing learned distributed representations in this way can provide some intuition regarding the quality of the embeddings and the structure of the learned space. In Figure \ref{fig:skipatomreduced}, the 200-dimensional learned SkipAtom vectors are plotted after utilizing t-SNE to reduce their dimensionality to 2. It is evident that there is a logical structure to the data. We see that the alkali metals are clustered together, as are the light non-metals, for example. The relative locations of the atoms in the plot reflect chemo-structural nuances gleaned from the dataset, and are not arbitrary.

To properly evaluate the quality of a learned distributed representation, they are utilized in the context of a task, and their performance compared to other representations. Here, we use the Elpasolite Formation Energy prediction task, and compare the performance of the SkipAtom vectors to the performance of other representations, namely, to Random vectors, One-hot vectors, Mat2Vec and Atom2Vec vectors. In the original study that introduced the task, atom vectors were 30- and 86-dimensional. We trained SkipAtom vectors with the same dimensions, and also with 200 dimensions, and evaluated them. The results are summarized in Table \ref{tab:elpasoliteformationenergy}.

\begin{table}[H]
\centering
\caption {Elpasolite Formation Energy prediction results after 10-fold cross-validation; mean best formation energy MAE on the test set after 200 epochs of training in each fold. Batch size was 32, learning rate was 0.001. Note that Dim refers to the dimensionality of the atom vector; the size of the input vector is 4 $\times$ Dim. All results were generated using the same procedure on identical train/test folds.}
\small
\begin{tabular}{ c c c }
\hline
\bf Representation & \bf Dim & \bf MAE (eV/atom) \\
\hline
Atom2Vec & 30 & 0.1477 $\pm$ 0.0078 \\
SkipAtom & 30 & 0.1183 $\pm$ 0.0050 \\
Random & 30 & 0.1701 $\pm$ 0.0081 \\
Atom2Vec & 86 & 0.1242 $\pm$ 0.0066 \\
One-hot & 86 & 0.1218 $\pm$ 0.0085 \\
SkipAtom & 86 & 0.1126 $\pm$ 0.0078 \\
Random & 86 & 0.1190 $\pm$ 0.0085 \\
Mat2Vec & 200 & 0.1126 $\pm$ 0.0058 \\
\bf SkipAtom & \bf 200 & \bf 0.1089 $\pm$ 0.0061 \\
Random & 200 & 0.1158 $\pm$ 0.0050 \\
\hline
\end{tabular}
\label{tab:elpasoliteformationenergy}
\end{table}

\begin{figure}[H]
\begin{center}
\includegraphics[scale=0.55]{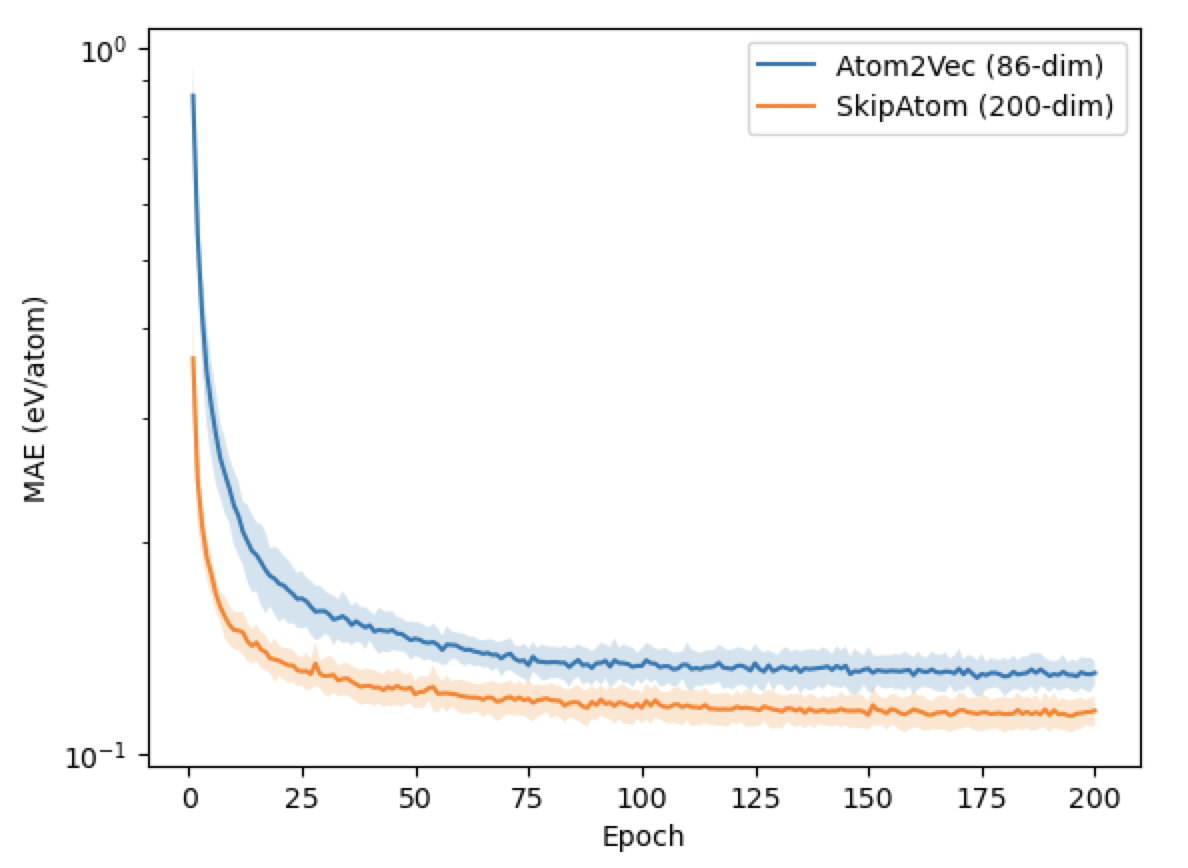}
\end{center}
\caption{A plot of the mean absolute error during training for the Elpasolite Formation Energy prediction task, for the Atom2Vec and SkipAtom representations. The average MAE over 10 folds is plotted.}
\label{fig:elpasolitetraining}
\end{figure}

For all embedding dimension sizes, SkipAtom outperforms the other representations on the Elpasolite Formation Energy task (Mat2Vec vectors were only available for this study in 200 dimensions, and Atom2Vec vectors, by virtue of how they are created, cannot have more dimensions than atom types represented). In Figure \ref{fig:elpasolitetraining}, a plot of how the mean absolute error changes during training demonstrates that the SkipAtom representation achieves better results from the beginning of training, and maintains the performance throughout.

\begin{figure}[H]
\begin{center}
\includegraphics[scale=0.285]{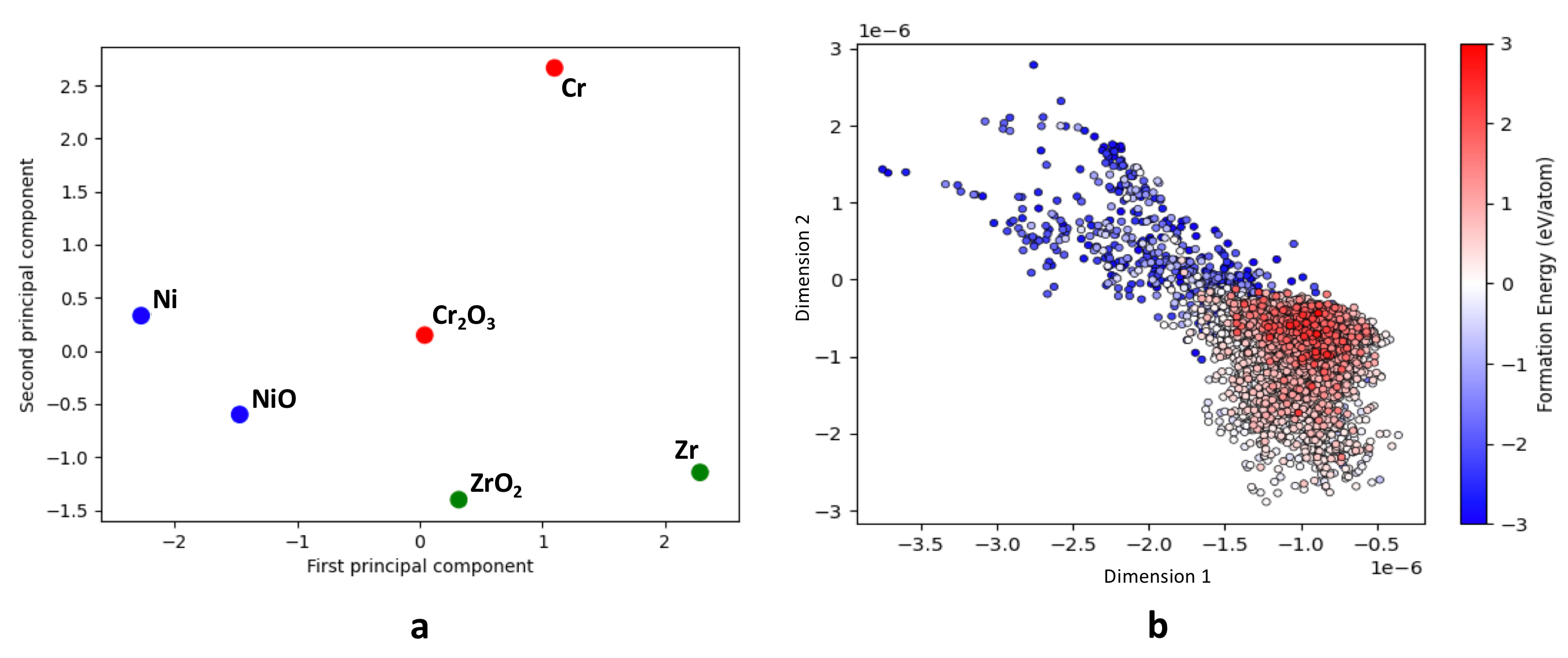}
\end{center}
\caption{\textbf{a} Plot of 200-dimensional SkipAtom vectors for Cr, Ni, and Zr, and their mean-pooled oxides, dimensionally reduced using PCA. \textbf{b} Plot of a sampling of the dimensionally-reduced compound vectors for the OQMD Dataset Formation Energy task, mapped to their associated physical values. The points are sum-pooled one-hot vectors reduced using t-SNE with a Hamming distance metric. The sum-pooled one-hot representation was the best performing for the task.}
\label{fig:tsne-boa-delta_e-hamming}
\end{figure}

Similar to atom vectors, compound vectors formed by the pooling of atom vectors can be dimensionally reduced, and visualized with t-SNE, or with PCA (Figure \ref{fig:tsne-boa-delta_e-hamming}a). In Figure \ref{fig:tsne-boa-delta_e-hamming}b, a sampling of several thousand compound vectors, formed by the sum-pooling of one-hot vectors, were reduced to 2 dimensions using t-SNE, and plotted. Additionally, since each compound vector represents a compound in the OQMD dataset, which contains associated formation energies, a color is assigned to each point in the plot denoting its formation energy. A clear distinction can be made across the spectrum of compounds and their formation energies. The vector representations derived from the composition of atom vectors appear to have preserved the relationship between atomic composition and formation energy.

Again, as with atom vectors, the quality of a compound vector is best established by comparing its performance in a task. To evaluate the quality of pooled atom vectors, 9 predictive tasks were utilized, as described in Table \ref{tab:tasktypes}. The performance on the benchmark regression tasks is summarized in Table \ref{tab:regressiontasks}, and the performance on the benchmark classification tasks is summarized in Table \ref{tab:classificationtasks}. Finally, the performance on the OQMD Formation Energy prediction task is summarized in Table \ref{tab:oqmdformationenergy}.

\begin{table}[H]
\caption {Benchmark regression task results after 2-repeated 5- or 10-fold cross-validation; mean best MAE on the test set after 100 epochs of training in each fold. All results were generated using the same procedure on identical train/test folds. TBG refers to the Theoretical Band Gap task (MAE in eV), BM to the Bulk Modulus task (MAE in log(GPa)), SM to the Shear Modulus task (MAE in log(GPa)), and RI to the Refractive Index task (MAE in $n$). These tasks make use of structure information. EBG refers to the Experimental Band Gap task (MAE in eV), and it makes use of composition only. Only the best results for each representation are reported. The pooling procedure varies between results; blue results represent sum-pooling, red results represent mean-pooling, and teal results represent max-pooling. Numbers in parentheses represent the standard deviation to one part in $10^{4}$. See the Supplementary Information tables for more detailed results.}
\footnotesize
\begin{tabular*}{\textwidth}{ c c c c c c c }
\hline
\bf Representation & \bf Dim & \bf EBG & \bf TBG & \bf BM & \bf SM & \bf RI \\
\hline
SkipAtom & 86 & \color{blue}0.3495\it(20) & \color{blue}0.2791\it(8) & \color{red}0.0789\it(2) & \color{blue}\bf 0.1014\it(1) & \color{red}0.3275\it(4) \\
Atom2Vec & 86 & \color{blue}0.3922\it(87) & \color{blue}0.2692\it(8) & \color{blue}0.0795\it(5) & \color{blue}0.1029\it(0) & \color{red}0.3308\it(16) \\
Bag-of-Atoms / One-hot & 86 & \color{blue}0.3797\it(22) & \color{blue}0.2611\it(8) & \color{blue}0.0861\it(2) & \color{blue}0.1137\it(5) & \color{blue}0.3576\it(2) \\
ElemNet / One-hot & 86 & \color{red}0.4060\it(72) & \color{red}\bf 0.2582\it(3) & \color{red}0.0853\it(1) & \color{red}0.1155\it(1) & \color{red}0.3409\it(16) \\
One-hot & 86 & \color{teal}0.3823\it(46) & \color{teal}0.2603\it(4) & \color{teal}0.0861\it(3) & \color{teal}0.1140\it(2) & \color{teal}0.3547\it(13) \\
Random & 86 & \color{blue}0.4109\it(58) & \color{red}0.3180\it(16) & \color{red}0.0908\it(4) & \color{blue}0.1195\it(2) & \color{red}0.3593\it(6) \\
Mat2Vec & 200 & \color{blue}0.3529\it(7) & \color{blue}0.2741\it(2) & \bf \color{blue}0.0776\it(0) & \color{blue}\bf 0.1014\it(2) & \color{red}\bf 0.3236\it(17) \\
SkipAtom & 200 & \color{blue}\bf 0.3487\it(85) & \color{blue}0.2736\it(8) & \color{red}0.0785\it(0) & \color{blue}\bf 0.1014\it(0) & \color{red}0.3247\it(15) \\
Random & 200 & \color{blue}0.4058\it(4) & \color{blue}0.3083\it(21) & \color{red}0.0871\it(1) & \color{red}0.1163\it(2) & \color{red}0.3543\it(6) \\
\hline
\end{tabular*}
\label{tab:regressiontasks}
\end{table}

\begin{table}[H]
\caption {Benchmark classification task results after 2-repeated 5-fold stratified cross-validation; mean best ROC-AUC on the test set after 100 epochs of training in each fold. All results were generated using the same procedure on identical train/test folds. TM refers to the Theoretical Metallicity task, and makes use of structure information. BMGF refers to the Bulk Metallic Glass Formation task, and EM to the Experimental Metallicity task. These last two do not make use of structure information. Only the best results for each representation are reported. The pooling procedure varies between results; blue results represent sum-pooling, red results represent mean-pooling, and teal results represent max-pooling. See the Supplementary Information tables for more detailed results.}
\footnotesize
\begin{tabular*}{\textwidth}{ c c c c c }
\hline
\bf Representation & \bf Dim & \bf TM & \bf BMGF & \bf EM \\
\hline
SkipAtom & 86 & \color{blue}0.9520 $\pm$ 0.0002 & \color{red}0.9436 $\pm$ 0.0010 & \color{blue}0.9645 $\pm$ 0.0012 \\
Atom2Vec & 86 & \color{blue}0.9526 $\pm$ 0.0001 & \color{red}0.9316 $\pm$ 0.0012 & \color{blue}0.9582 $\pm$ 0.0008 \\
Bag-of-Atoms / One-hot & 86 & \color{blue}0.9490 $\pm$ 0.0002 & \color{blue}0.9277 $\pm$ 0.0004 & \color{blue}0.9600 $\pm$ 0.0012 \\
ElemNet / One-hot & 86 & \color{red}0.9477 $\pm$ 0.0001 & \color{red}0.9322 $\pm$ 0.0014 & \color{red}0.9485 $\pm$ 0.0007 \\
One-hot & 86 & \color{teal}0.9487 $\pm$ 0.0003 & \color{teal}0.9289 $\pm$ 0.0016 & \color{teal}0.9599 $\pm$ 0.0014 \\
Random & 86 & \color{blue}0.9444 $\pm$ 0.0000 & \color{red}0.9274 $\pm$ 0.0006 & \color{blue}0.9559 $\pm$ 0.0021 \\
Mat2Vec & 200 & \color{blue}\bf 0.9528 $\pm$ 0.0002 & \color{red}0.9348 $\pm$ 0.0024 & \color{blue}\bf 0.9655 $\pm$ 0.0014 \\
SkipAtom & 200 & \color{blue}0.9524 $\pm$ 0.0001 & \bf \color{red}0.9349 $\pm$ 0.0019 & \color{blue}0.9645 $\pm$ 0.0008 \\
Random & 200 & \color{blue}0.9453 $\pm$ 0.0001 & \color{red}0.9302 $\pm$ 0.0016 & \color{blue}0.9541 $\pm$ 0.0002 \\
\hline
\end{tabular*}
\label{tab:classificationtasks}
\end{table}

\begin{table}[H]
\centering
\caption {OQMD Dataset Formation Energy prediction results after 10-fold cross-validation; mean best formation energy MAE on the test set after 100 epochs of training in each fold. All results were generated using the same procedure on identical train/test folds.}
\small
\begin{tabular}{ c c c c }
\hline
\bf Representation & \bf Dim & \bf Pooling & \bf MAE (eV/atom) \\
\hline
SkipAtom & 86 & sum & 0.0420 $\pm$ 0.0005 \\
Atom2Vec & 86 & sum & 0.0396 $\pm$ 0.0004 \\
\bf Bag-of-Atoms / One-hot & \bf 86 & \bf sum & \bf 0.0388 $\pm$ 0.0002 \\
ElemNet / One-hot & 86 & mean & 0.0427 $\pm$ 0.0007 \\
Random & 86 & sum & 0.0440 $\pm$ 0.0004 \\
Mat2Vec & 200 & sum & 0.0401 $\pm$ 0.0004 \\
SkipAtom & 200 & sum & 0.0408 $\pm$ 0.0003 \\
Random & 200 & sum & 0.0417 $\pm$ 0.0004 \\
\hline
\end{tabular}
\label{tab:oqmdformationenergy}
\end{table}

In the benchmark regression and classification task results, there isn't a clear atom vector or pooling method that dominates. The 200-dimensional representations generally appear to perform better than the smaller 86-dimensional representations. Though not evident from Tables \ref{tab:regressiontasks} and \ref{tab:classificationtasks}, sum- and mean-pooling outperform max-pooling (see Supplementary Information tables). The pooled Mat2Vec representations are notable, in that they achieve the best results in 4 of the 8 benchmark tasks, while pooled SkipAtom representations are best in 2 of the 8 benchmark tasks. Pooled Random vectors tend to under-perform, though not always by a very large margin. This may not be so surprising, since random vectors exhibit quasi-orthogonality as their dimensionality increases, and thus may have the same functional characteristics as one-hot vectors. \cite{kainen2020quasiorthogonal} On the OQMD Formation Energy prediction task, the Bag-of-Atoms representation yields the best results, significantly outperforming both the distributed representations, and the mean-pooled one-hot representation originally used in the ElemNet paper, that introduced the task.

\begin{figure}[H]
\begin{center}
\includegraphics[scale=0.38]{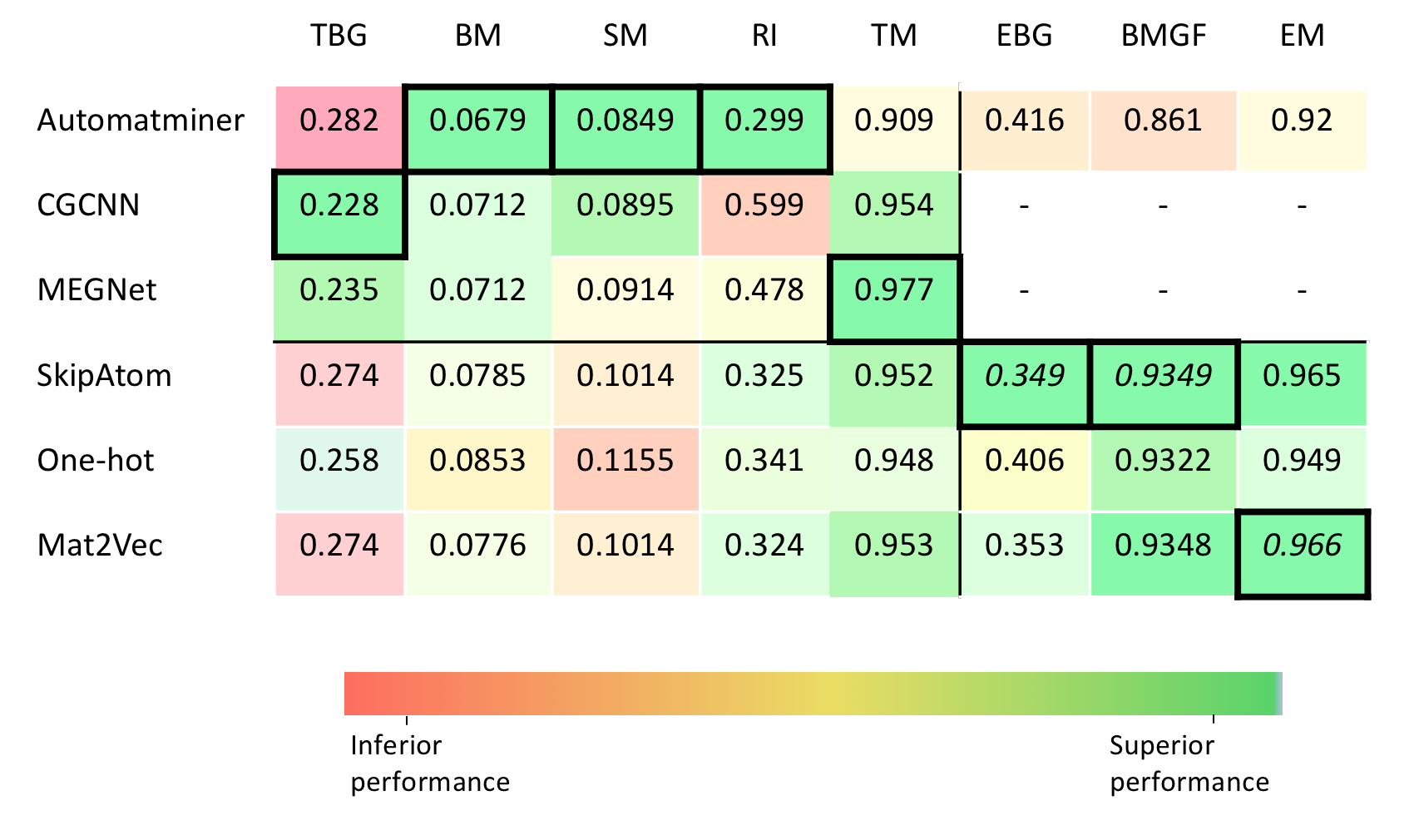}
\end{center}
\caption{A comparison between the results of the methods described in the current work and existing state-of-the-art results on benchmark tasks. TBG refers to the Theoretical Band Gap task (MAE in eV), BM to the Bulk Modulus task (MAE in log(GPa)), SM to the Shear Modulus task (MAE in log(GPa)), RI to the Refractive Index task (MAE in $n$), and TM to the Theoretical Metallicity task (ROC-AUC). These tasks make use of structure information. EBG refers to the Experimental Band Gap task (MAE in eV), BMGF to the Bulk Metallic Glass Formation task (ROC-AUC), EM to the Experimental Metallicity task (ROC-AUC). These tasks make use of composition only. The results that are outlined in bold represent the best score for that task. Italicized results represent a new state-of-the-art result. As described in the Protocols section of this report, the same methodology was used to obtain the results for all of the algorithms in the table.}
\label{fig:automatminer}
\end{figure}

A noteworthy aspect of these results is how the pooled atom vector representations compare to the published state-of-the-art values on the 8 benchmark tasks from the Matbench test suite. Figure \ref{fig:automatminer} depicts this comparison. Indeed, the models described in this report outperform the existing benchmarks on tasks where only composition is available (namely, the Experimental Band Gap, Bulk Metallic Glass Formation, and Experimental Metallicity tasks), and represent new state-of-the-art results. Also, on the Theoretical Metallicity task and the Refractive Index task, the pooled SkipAtom, Mat2Vec and one-hot vector representations perform comparably, despite making use of composition information only.

\section{Conclusions}

NLP researchers have learned many lessons regarding the computational representations of words and sentences. It could be fruitful for computational materials scientists to borrow techniques from the study of Computational Linguistics. Above, we have described how making an analogy between words and sentences, and atoms and compounds, allowed us to borrow both a means of learning atom representations, and a means of forming compound representations by pooling operations on atom vectors. Consequently, we draw the following conclusions: i) effective computational descriptors of atoms can be derived from freely available and growing materials databases; ii) effective computational descriptors of compounds can be easily constructed by straightforward pooling operations of the atom vectors of the constituent atoms; iii) properties of materials can often be accurately predicted even where only chemical composition information is available. 

The SkipAtom representation can be derived from a dataset of readily accessible compound structures. Moreover, the training process is lightweight enough that it can be performed on a good quality laptop on a scale of minutes to several hours (given the atom pairs). This highlights some important differences between SkipAtom and Atom2Vec and Mat2Vec. Training of the Mat2Vec representation requires the curation of millions of journal abstracts, and a subsequent classification step for retaining only the most relevant abstracts. Additionally, pre-processing of the tokens in the text must be carried out to identify valid chemical formulae through the use of custom rules and regular expressions. On the other hand, since SkipAtom makes direct use of the information in materials databases, no special pre-processing of the chemical information is required. Although the procedures for creating Mat2Vec and SkipAtom vectors have been incorporated into publicly available software libraries, the conceptually simpler SkipAtom approach leaves little room for ambiguity that might result from manually written chemical information extraction rules. When compared to Atom2Vec, a principal difference is that SkipAtom vectors are not limited in size by the number of atom types available. This allows larger SkipAtom vectors to be trained, and, as is evident from the results described above, larger vectors generally perform better on tasks. Overall, we believe SkipAtom is a more accessible tool for computational materials scientists, allowing them to readily train expressive atom vectors on chemical databases of their choosing, and to take advantage of the growing information in these databases over time.

The ElemNet architecture demonstrated that the incorporation of composition information alone could result in good performance when predicting chemical properties. In this work, we have extended the result, and shown how such an approach performs in a variety of different tasks. Perhaps surprisingly, the combination of a deep feed-forward neural network with compound representations consisting of composition information alone results in competitive performance when comparing to approaches that make use of structural information. We believe this is a valuable insight, since high-throughput screening endeavours, in the search for new materials with desired properties, often target areas of chemical space where only composition is known. We envision performing large sweeps of chemical space, in relatively shorter periods of time, since structural characteristics of the compounds would not need to be computed, and only composition would be used. The results presented here could motivate more extensive and computationally cheaper screening.

Going forward, there are a number of different avenues that can be explored. First, the atom vectors generated using the SkipAtom approach can be explored in different contexts, such as in combination with structural information. For example, graph neural networks, such as the MEGNet architecture \cite{chen2019graph}, can accept as input any atom representation one chooses. It would be interesting to see if starting with pre-trained SkipAtom vectors could improve the performance of these models, where structure information is also incorporated. Alternatively, chemical compound vectors formed by pooling SkipAtom vectors can be directly concatenated with vectors that contain structure information, thus complementing the pooled atom vectors with more information. A candidate for encoding structure information is the Coulomb Matrix (in vectorized form), a descriptor which encodes the electrostatic interactions between atomic nuclei. \cite{rupp2012fast} Finally, one limitation of the SkipAtom approach is that it does not provide representations of atoms in different oxidation states. Since it is (often) possible to unambiguously infer the oxidation states of atoms in compounds, it is, in principle, possible to construct a SkipAtom training set of pairs of atoms in different oxidation states. The number of atom types would increase by several fold, but would still be within limits that allow for efficient training. It would be interesting to explore the results of forming compound representations using such vectors for atoms in various oxidation states.

\section{Code Availability}

The code for creating and using the SkipAtom vectors is open source, released under the GNU General Public License v3.0. The code repository is accessible online, at:\\
\href{https://github.com/lantunes/skipatom}{\color{blue}{https://github.com/lantunes/skipatom}}\\
The repository also contains pre-trained 200-dimensional SkipAtom vectors for 86 atom types that can be immediately used in materials informatics projects.

\section{Data Availability}

The data that support the findings of this study are available as follows: The materials data that was used to learn the SkipAtom embeddings are publicly available online at \href{https://materialsproject.org/}{\color{blue}{https://materialsproject.org/}}. The elpasolite formation energy training data are publicly available online at \href{https://journals.aps.org/prl/abstract/10.1103/PhysRevLett.117.135502}{\color{blue}{https://journals.aps.org/prl/abstract/10.1103/PhysRevLett.117.135502}}, in the Supplemental Material section. The datasets comprising the Matbench tasks are publicly available at \href{https://hackingmaterials.lbl.gov/automatminer/datasets.html}{\color{blue}{https://hackingmaterials.lbl.gov/automatminer/datasets.html}}. The Mat2Vec pre-trained embeddings are publicly available online and can be downloaded by following the instructions at \href{https://github.com/materialsintelligence/mat2vec}{\color{blue}{https://github.com/materialsintelligence/mat2vec}}. The Atom2Vec embeddings are publicly available online and can be obtained from\\ 
\href{https://github.com/idocx/Atom2Vec}{\color{blue}{https://github.com/idocx/Atom2Vec}}.

\section{Author Contributions}

L.M.A. conceived the project, designed and performed the experiments, and drafted the manuscript. R.G.-C. and K.T.B. supervised and guided the project. All authors reviewed, edited and approved the manuscript.

\section{Competing Interests}

The authors declare no competing interests.

\bibliographystyle{naturemag}
\bibliography{references}

\end{document}